\documentclass[%
 reprint,
 amsmath,amssymb,
 aps,
 longbibliography,
pre,
]{revtex4-2}

\usepackage[caption=false]{subfig}
\usepackage{graphicx}
\usepackage{dcolumn}
\usepackage{bm}
\usepackage{xcolor}
\usepackage{comment}
\usepackage {hyperref}
\hypersetup{
    colorlinks = true,
    linkcolor = blue,
    anchorcolor = blue,
    citecolor = blue,
    filecolor = blue,
    urlcolor = blue
    }
\DeclareMathAlphabet{\mathpzc}{OT1}{pzc}{m}{it}

\begin{document}

\preprint{APS/123-QED}

\title{Accurate Formula for the Effective Conductivity of Highly Clustered Two-Phase Materials}

\author{Murray Skolnick}
\affiliation{
    Department of Chemistry, Princeton University, Princeton, New Jersey 08544, USA
}%

\author{Salvatore Torquato}
\email{torquato@electron.princeton.edu}
\affiliation{
    Department of Chemistry, Department of Physics, Princeton Materials Institute, and Program in Applied and Computational Mathematics, Princeton University, Princeton New Jersey 08544, USA
}%

\date{\today}

\begin{abstract}

Two-phase heterogeneous materials arising in a variety of natural and synthetic situations exhibit a wide-variety of microstructures and thus display a broad-spectrum effective physical properties. Given that such properties of disordered materials generally depend on an infinite-set of microstructural correlation functions that are typically unobtainable, in practice, one must consider rigorous bounds and approximations on the effective properties that depend on a limited set of nontrivial microstructural information. Torquato derived [\textit{Random Heterogeneous Materials, Microstructure and Macroscopic Properties} (2002)] such an approximation for the effective thermal/electrical conductivity of two-phase microstructures that perturb about those that realize the well-known self-consistent formula; a key feature of such microstructures being phase-inversion symmetry which can be observed in certain cellular, bicontinuous, and porous materials. Here, we show via extensive numerical simulations that this virtually unexamined approximation, which depends on up to three-point correlations, predicts accurately the effective conductivities of various two- and three-dimensional two-phase microstructures across phase volume fractions in which both phases can form large and well-connected clusters; including certain phase-inversion symmetric microstructures as well as phase-inversion asymmetric dispersions of certain fully penetrable particles. We also highlight the accurate sensitivity of this approximation to phase-connectedness properties by showing that it predicts percolation thresholds that are in good quantitative and/or qualitative agreement with known thresholds for the models considered here. Finally, we discuss how this formula can be used to design materials with desirable effective conductivity properties and thus aid in materials by design.

\end{abstract}

\maketitle

\section{Introduction}\label{sec:introduction}

\begin{figure}
    \subfloat[\label{fig:labyrinth}]{\includegraphics[width=0.5\columnwidth]{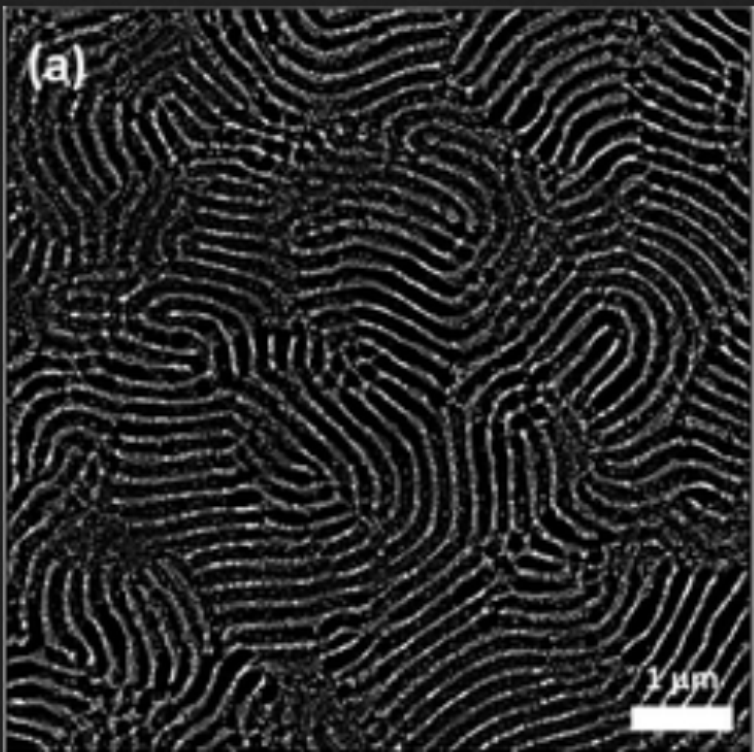}}~
    \subfloat[\label{fig:spinodal}]{\includegraphics[width=0.5\columnwidth]{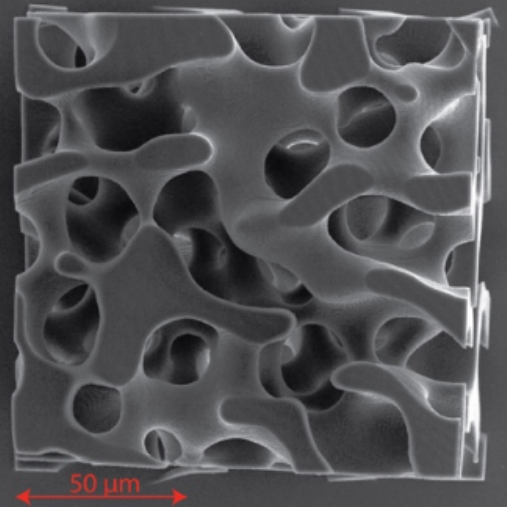}}
    \caption{Images of real material microstructures in which both phases form well-connected clusters and likely possess phase-inversion symmetry: (a) A scanning electron microscope (SEM) image of a labyrinth of silicon nano walls (light gray) and void (black) manufactured via solvent vapor annealing using a self-assembled block copolymer template \cite{Se21}. This particular system likely has silicon phase volume fraction $\phi_2\approx0.5$. (b) An SEM image of a spinodal architected metamaterial manufactured using two-photon polymerization Direct Laser Writing which has solid phase-volume fraction $\phi_2=0.30$ \cite{Hs19}. The microstructure shown in (a) appears to be phase-inversion symmetric since the geometries and topologies of its silicon and void phases are qualitatively similar to one another. The two-phase microstructure shown in (b) is likely phase-inversion symmetric since it was derived from a level cut of a concentration field simulated via the Cahn-Hilliard equation \cite{Hs19}, which treats the interactions between and dynamics of the separating phases identically \cite{Ca58,Ca65}. The image in (a) is licensed under Creative Commons BY 4.0. The image in (b) is reprinted from Ref. \cite{Hs19} with permission from Elsevier.}
    \label{fig:examples}
\end{figure}

Disordered heterogeneous two-phase materials in $d$-dimensional Euclidean space $\mathbb{R}^d$ abound in natural and synthetic situations; examples include composites, polymer blends, colloids, porous media, biological media, and complex fluids \cite{To02a, Sa03, Mi02, Pa16, To98b, Ch17, Ch18, Hu21}. Among the broad-spectrum of realizable microstructure phase geometries and topologies, the microstructures of certain two-phase materials like aerogels \cite{Brink90, Ro97}, interpenetrating cermets \cite{Po90, To99a, Ba22} and those derived from triply-periodic minimal surfaces \cite{Ju05, Zh21c, To24} consist of \textit{large and well-connected clusters of both phases.} Examples of real material microstructures with such phase geometries and topologies are the labyrinth of silicon nano walls \cite{Se21} shown in Fig. \ref{fig:labyrinth}, and the spinodal architected metamaterial \cite{Hs19} shown in Fig. \ref{fig:spinodal}. 

In general, the effective physical properties of heterogeneous materials depend on an infinite set of correlation functions \cite{Se89, To97b, To02a, Ki20, To21a}. For example, there is the countably infinite set of the standard $n$-point correlation functions $S^{(i)}_n(\mathbf{x}_1,...,\mathbf{x}_n)$ which give the probability of finding the points $\mathbf{x}_1,...,\mathbf{x}_n$ in phase $i=1,2$ for two-phase materials \cite{To02a}. As such infinite sets of correlation functions are typically unobtainable in practice, the capacity to predict effective material properties using theoretical techniques that involve limited sets of nontrivial microstructural information or ways of approximating higher order $n$-point information enhances our fundamental understanding of structure-property relationships and informs the design of structural and functional materials \cite{To02a, Mi02, Ki20a, To21a}. One such theoretical technique for predicting effective properties (i.e., conductivity and elastic moduli) is the \textit{strong-contrast} expansion whose terms consist of functionals of certain field quantities and $S^{(i)}_n(\mathbf{x}_1,...,\mathbf{x}_n)$ for all $n$ \cite{To02a, To85f, Se89, To97c, Br55}. Importantly, by utilizing an expansion parameter that is a \textit{rational} function of the phase properties, rather than simple differences of them, the strong-contrast expansions possess superior convergence properties for any phase property contrast ratio. Moreover, lower-order truncation of this rapidly converging series results in accurate approximations on the effective conductivity and elastic moduli for various classes of microstructures as we will now describe. 

\begin{figure}
    \subfloat[\label{fig:dispersion}]{\includegraphics[width=0.5\columnwidth]{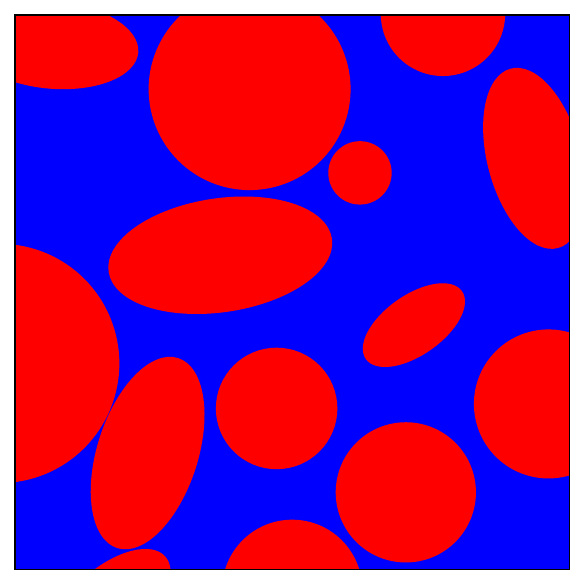}}~
    \subfloat[\label{fig:droplet}]{\includegraphics[width=0.5\columnwidth]{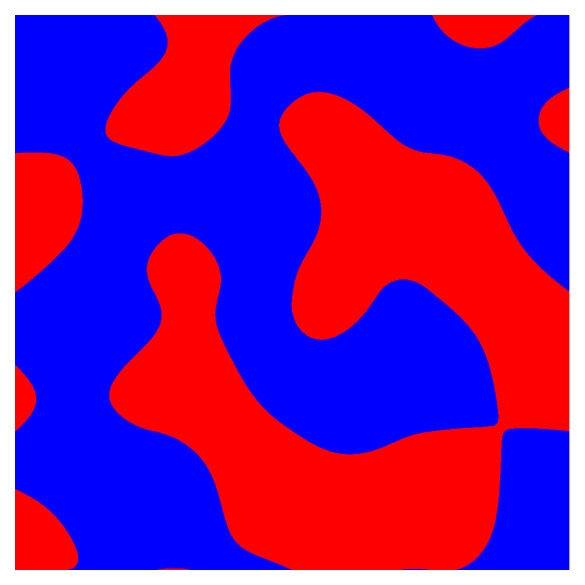}}
    \caption{Schematics illustrating the distinct classes of microstructures for which the different three-point strong-contrast approximations \eqref{eqn:20.83} and \eqref{eqn:20.84} are accurate. The dispersion shown in (a) consists of non-overlapping particles (red) embedded in a matrix (blue), while the microstructure shown in (b) consists of amorphous inclusions of both the red and blue phases. Approximation \eqref{eqn:20.83} is highly accurate for microstructures like that shown in (a) in which one phase does do not form large clusters. By contrast, as we will show in this paper, the approximation \eqref{eqn:20.84} is accurate for microstructures resembling that shown in (b) in which both phases can form large clusters.}
    \label{fig:schematics}
\end{figure}

As the aforementioned approximations incorporate microstructural information up to the level of the three-point correlation function $S_3$, they are sometimes referred to as the ``three-point approximations" \cite{To02a}. For example, the three-point approximation for the effective conductivity $\sigma_e$ for dispersions of particles in a matrix is given by \cite{To85f} 
\begin{equation}
    \frac{\sigma_e}{\sigma_q} = \frac{\left(1 + (d-1)\phi_p\beta_{pq} - (d-1)\phi_1\zeta_p\beta_{pq}^2\right)}{1 - \phi_p\beta_{pq} - (d-1)\phi_q\zeta_p\beta_{pq}^2},\label{eqn:20.83}
\end{equation}
in which
\begin{equation}
    \beta_{pq}=\frac{\sigma_p-\sigma_q}{\sigma_p + (d-1)\sigma_q}\label{eqn:beta_pq}
\end{equation}
is the scalar ``polarizability", $\sigma_p$ ($\sigma_q$) is the conductivity of phase $p$ ($q$), and the three-point parameter $\zeta_p$ is a functional over $S_1$, $S_2$, and $S_3$ (see Sec. \ref{sec:zeta} for additional details). Approximation \eqref{eqn:20.83} is highly accurate for dispersions in which the particles (phase 2), generally, do not form large clusters \cite{To85f, Se89, To97c, To98c, To02a}. A schematic image illustrating the class of dispersions for which approximation \eqref{eqn:20.83} is accurate is shown in Fig. \ref{fig:dispersion}. Formula \eqref{eqn:20.83} has been shown to predict accurately the effective thermal/electrical conductivities of various disordered packings of disks and spheres \cite{Gi14,Sk24c}, random packings of ellipsoids and Platonic solids \cite{Gi15}, and by mathematical analogy, the effective diffusion coefficients of certain chromatography columns whose microstructures consist of disordered packings of spherical beads \cite{Gr11b, Gr11c, De22}. 

Torquato derived a different, virtually unexplored, three-point strong-contrast approximation for the effective conductivity that was designed to predict $\sigma_e$ for a structurally distinct class of two-phase material microstructures in which both phases can form large and well-connected clusters, including \textit{bicontinuous} \footnote{Bicontinuous microstructures are those in which both phases are topologically connected across the sample. This topological feature (i.e., percolation of both phases) is common in 3D systems, but is rare in 2D ones \cite{To02a}.} systems, across volume fractions, which is given by
\begin{multline}
    \phi_2 \frac{\sigma_e + (d-1)\sigma_1}{\sigma_e-\sigma_1}  + \phi_1 \frac{\sigma_e + (d-1)\sigma_2}{\sigma_e-\sigma_2} = \\ 
    2 - d - (d-1)\beta_{21}\left[\phi_1\zeta_2 - \frac{\phi_2(1-\zeta_2)}{1 + (d-2)\beta_{21}}\right].\label{eqn:20.84}
\end{multline}
A schematic image illustrating the class of highly clustered microstructures for which approximation \eqref{eqn:20.84} can be applied is shown in Fig. \ref{fig:droplet}. In this paper, we examine for the first time the accuracy of this approximation using comprehensive numerical simulations. Torquato predicted that this formula would be accurate for microstructures that resemble those that realize the well-known ``self-consistent" formula, a key feature of which being phase-inversion symmetry \cite{To02a}. In such phase-inversion symmetric microstructures, like the nano wall labyrinth and spinodal architected metamaterial microstructures shown in Fig. \ref{fig:examples}, the morphology of phase 1 at volume fraction $\phi_1$ is statistically identical that of phase 2 in the system where the volume fraction of phase 1 is $1 - \phi_1$ \cite{To02a} (see Sec. \ref{sec:models} for additional details). 

Beyond being phase-inversion symmetric, we note that microstructures that realize the self-consistent formula also have a nontrivial percolation threshold (i.e., the critical volume fraction at which phase 2 forms a sample-spanning cluster) $\phi_{2c}=1/d$ and will be bicontinuous in 3D for $1/3<\phi_2<2/3$ \cite{To02a}. Since microstructures with percolation thresholds at intermediate volume fractions necessarily possess large clusters away but near the thresholds, we anticipate that formula \eqref{eqn:20.84} should also be accurate for highly clustered two-phase materials, phase-inversion symmetric or not, which we verify below. Note that Torquato's complete derivation of \eqref{eqn:20.84} using strong-contrast expansions as well as a detailed discussion on how it perturbs about the self-consistent formula are provided in Ref. \cite{To02a}. Henceforth, we refer to approximation \eqref{eqn:20.84} as the ``self-consistent strong-contrast approximation" which, by mathematical analogy, can also be used to compute the effective diffusion coefficient $D_e$, dielectric constant $\epsilon_e$, and the magnetic permeability $\mu_e$ of two-phase materials \cite{To02a}.

We demonstrate the high predictive power of the self-consistent strong-contrast approximation \eqref{eqn:20.84} by comparing its predictions for $\sigma_e$ to those obtained via simulations for several physically realistic disordered model 2D and 3D microstructures that exhibit phase clustering across volume fractions. We specifically consider phase-inversion symmetric ``Debye random media" \cite{De57, Ye98a} and symmetric-cell materials (see Sec. \ref{sec:zeta} for definitions) with square and cubical cells \cite{To02a}, as well as phase-inversion asymmetric dispersions of fully penetrable disks, spheres, elongated ellipses, and oblate ellipsoids \cite{To02a}. Beyond accurate predictions for $\sigma_e$, we also derive from \eqref{eqn:20.84} non-trivial predictions for $\phi_{2c}$ that depend on microstructural information via the parameter $\zeta_2$. We show that these predicted percolation thresholds are in good qualitative and/or quantitative agreement with known values of $\phi_{2c}$ for the models considered here. We further highlight the sensitivity of \eqref{eqn:20.84} to microstructure percolation properties by examining it as a function of phase conductivity contrast ratio $\sigma_2/\sigma_1$ across all volume fractions. 

The rest of the paper is organized as follows: information about the three-point parameter $\zeta_p$ is provided in Sec. \ref{sec:zeta}. The model microstructures, their values of the three-point parameters across volume fractions, and phase-inversion symmetry/asymmetry are summarized in Sec. \ref{sec:models}. The main results are presented in Sec. \ref{sec:results}, and conclusions and future directions are discussed in Sec. \ref{sec:conclusion}. 

\section{The Three-Point Parameter $\zeta_p$}\label{sec:zeta}

Here, we summarize key details about the three-point microstructural parameter $\zeta_p$ that appears in approximations \eqref{eqn:20.83} and \eqref{eqn:20.84}. We note that $\zeta_p$ also arises in rigorous bounds on the effective conductivity and the elastic moduli  that improve upon the well-known Hashin-Shtrikman bounds \cite{Be66, Mc70, Si72, Sc76a, To80, Mi81b, Mi82a, Mi82b, Ph82, Mi84, To85f, Gi95a, Gi95b}, strong-contrast expansions truncated after third order for the effective elastic moduli \cite{To02a, To97c}, as well as an approximation for the fluid permeability of certain materials \cite{To20, Kl21b, To24}. The parameter $\zeta_p$ for phase $p$ of a two-phase microstructure is specifically given by \begin{multline}
    \zeta_p = \frac{4}{\pi\phi_1\phi_2}\int_0^{\infty}\frac{dr}{r}\int_0^{\infty}\frac{ds}{s}\int_0^{\pi} d\theta \cos(2\theta) \\
    \left[S_3^{(p)}(r,s,\theta) - \frac{S_2^{(p)}(r)S_2^{(p)}(s)}{S_1^{(p)}}\right],\label{eqn:zeta2D}
\end{multline}
for 2D systems, and by
\begin{multline}
    \zeta_p = \frac{9}{2\phi_1\phi_2}\int_0^{\infty}\frac{dr}{r}\int_0^{\infty}\frac{ds}{s}\int_{-1}^{1} d(\cos\theta) P_2(\cos\theta) \\
    \left[S_3^{(p)}(r,s,\theta) - \frac{S_2^{(p)}(r)S_2^{(p)}(s)}{S_1^{(p)}}\right].\label{eqn:zeta3D}
\end{multline} 
for 3D ones \cite{To85}. The quantities $S^{(p)}_1$, $S_2^{(p)}(r)$, and $S_3^{(p)}(r,s,\theta)$ are, respectively, the probabilities of finding in phase $p$ a single point, the end points of a line segment of length $r$, and the vertices of a triangle with two sides of length $r$, $s$ and angle $\theta$ between them. In integral \eqref{eqn:zeta3D}, $P_2$ is the Legendre polynomial of order 2. Note that $S_1^{(p)}$ is simply the volume fraction of phase $p$, $\phi_p$, and that $0\leq\zeta_p\leq1$ and $\zeta_1 + \zeta_2=1$. As the three-point correlation function $S_3^{(p)}$ is generally not known analytically for most microstructures \cite{To85b, To85c, To85d, To85e, To85f, Sm89, To02a}, the parameter $\zeta_p$ must be ascertained directly from microstructural samples using numerical methods. We recently developed such a method for computing $\zeta_p$ that improves upon the accuracy and computational performance of previous methods \cite{Sk24c} which we employ in this paper as described in Sec. \ref{sec:threepoint_results}. 

In a forthcoming manuscript, we show via comprehensive numerical simulations that $\zeta_2$ tends to be larger for microstructures in which phase 2 is more well-connected and ramified for a given value of $\phi_2$ \cite{Sk24c}. symmetric-cell materials, which are constructed by partitioning space into cells of arbitrary size and shape and then randomly designating them as phases 1 and 2 with respective probabilities $\phi_1$ and $\phi_2$ \cite{Mi69, To02a}, specifically have
\begin{equation}
    \zeta_2 = \phi_2 + (1 - 2\phi_2)g,\label{eqn:z2_scm}
\end{equation}
in which $0\leq g\leq 1$ is a parameter that depends on the cell-shape, and is thus referred to as the ``cell-shape parameter." Specifically, $g$ is a functional over a certain three-point statistical quantity associated with the cell geometry \cite{To02a}. By virtue of their construction, all symmetric-cell materials are phase-inversion symmetric \cite{To02a}. One can show that such materials with circular/spherical cells have shape parameter $g=0$, while those with needle-like cells in 2D and disk-shaped cells in 3D have $g=1$ \cite{To02a}. We note that space can be tessellated by polydisperse circles/spheres or ellipses/ellipsoids which necessarily must have a continuous range of sizes ranging to the infinitesimally small. Moreover, Kim and Torquato recently showed that such multi-scale tessellations of disks and spheres can be effectively realized computationally using a tessellation-based procedure \cite{Ki19a,Ki19b}. Finally, Eq. \eqref{eqn:z2_scm} implies that all symmetric-cell materials have $\zeta_2=1/2$ for $\phi_2=1/2$ which is a general property of \textit{all} phase-inversion symmetric microstructures for reasons detailed in Ref. \cite{To02a}. 

\section{Model Microstructures and Three-Point Parameters}\label{sec:models}

In this section, we summarize the salient details of the four model 2D and 3D microstructures that we use as benchmarks to demonstrate the accuracy of the self-consistent strong-contrast approximation \eqref{eqn:20.84}. By virtue of their construction, these microstructures all exhibit much greater degrees of phase clustering than that observed in dispersions of well-separated particles in a matrix for a given volume fraction. We also compute their respective three-point parameters $\zeta_2$ across volume fractions and show how one can extract cell-shape parameters $g$ from these data for Debye random media and thus interpret these microstructures as ``effective" symmetric-cell materials.

\subsection{Models}

We specifically consider symmetric-cell materials with square and cubical cells with side length $a$. These microstructures have two-point correlation functions $S^{(p)}_2(r)=\phi_p\phi_q W_2(r;a) + \phi_p^2$ in which $W_2(r;a)$ is the ``single-cell weight" function for which exact expressions are given in Ref. \cite{To02a}. As these microstructures are sometimes referred to as ``random checkerboards", we use the acronym RCB to refer to them henceforth. Here, the RCB microstructures serve as a convenient platform to test the accuracy of \eqref{eqn:20.84} for symmetric-cell microstructures as they are single scale and thus generating realizations of them and ascertaining their effective conductivities is computationally straightforward. We also examine so-called ``Debye random media" (DRM) which are phase-inversion symmetric \cite{Ma20b,Sk21} models of two-phase media that are defined entirely by the two-point correlation function $S^{(p)}_2(r)=\phi_p\phi_q\exp(-r/a) + \phi_p^2$ in which $a>0$ is the characteristic length scale of the microstructure. Such an exponential form of $S_2(r)$ was hypothesized by Debye et al. to model 3D microstructures with phases of ``fully random shape, size, and distribution" \cite{De57}, and were shown to be realizable in 2D \cite{Ye98a} and 3D \cite{To20}. Note that DRM have been shown to be a good approximation of certain microstructures, including Fontainebleau sandstone \cite{Co96}, snow \cite{Va20}, and certain mesoporous materials \cite{Go18}. 

\begin{figure}
    \subfloat[\label{fig:rcb}]{\includegraphics[width=0.20\textwidth]{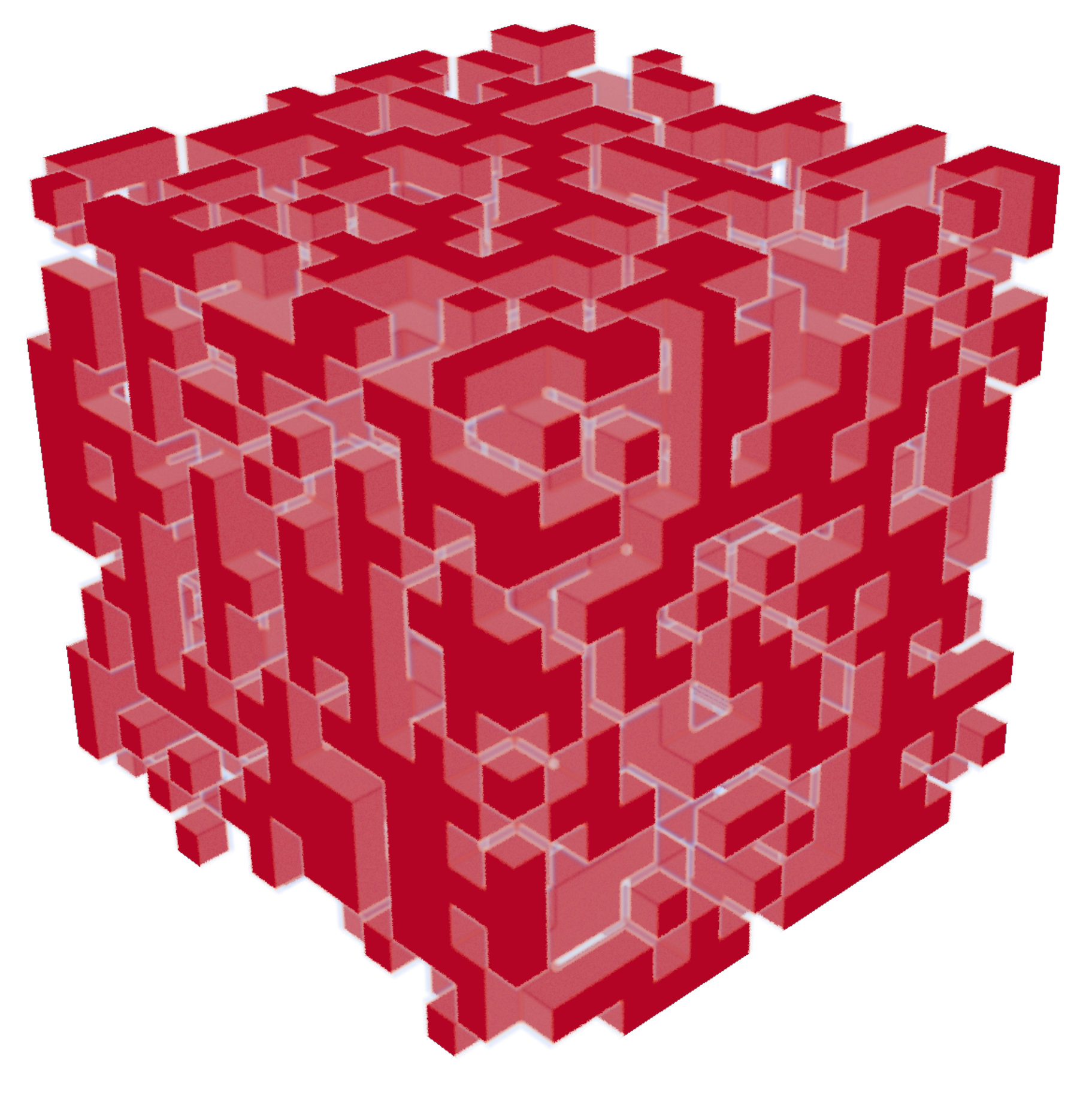}}~
    \subfloat[\label{fig:drm}]{\includegraphics[width=0.20\textwidth]{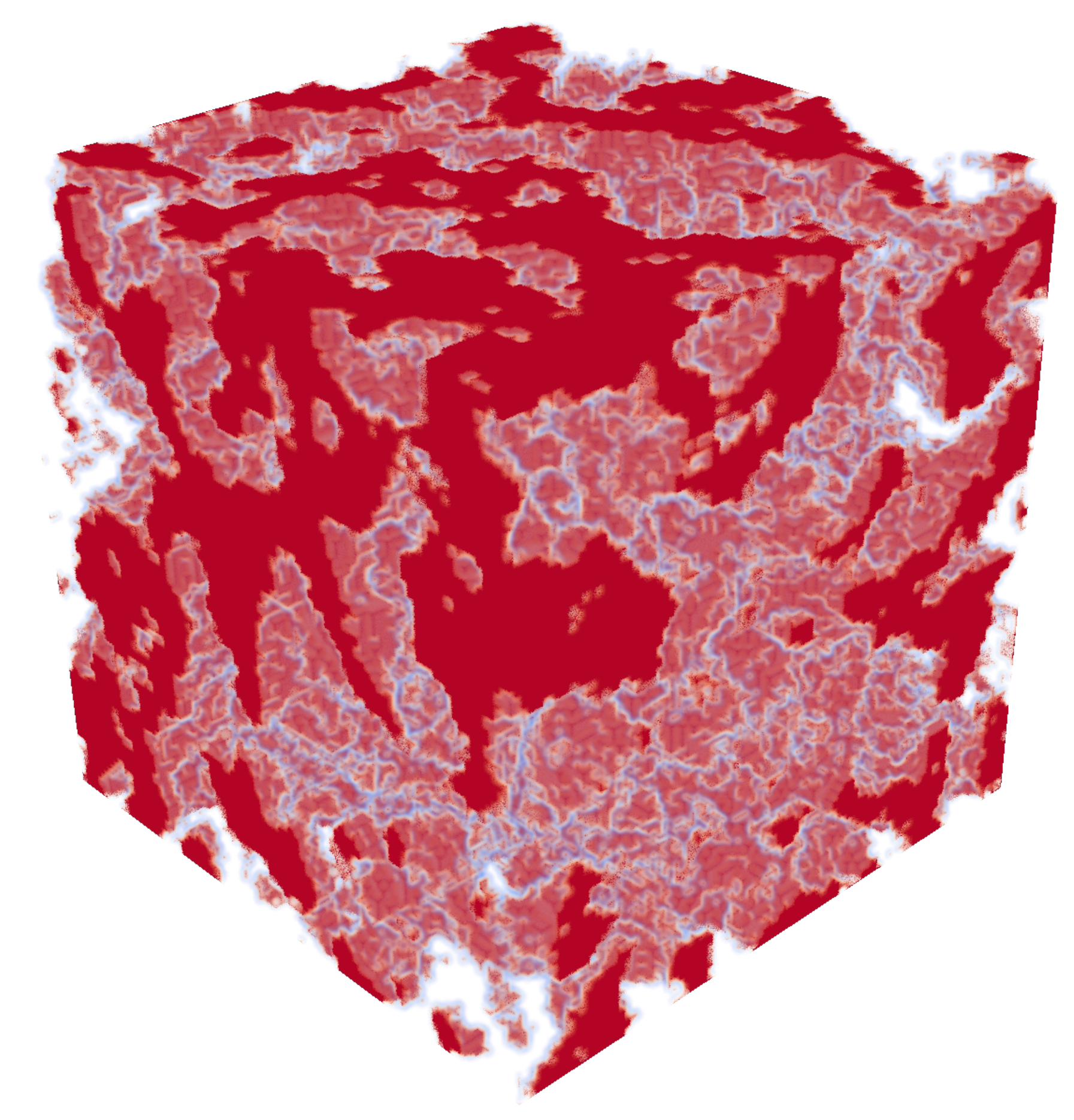}}\\
    \subfloat[\label{fig:fps}]{\includegraphics[width=0.20\textwidth]{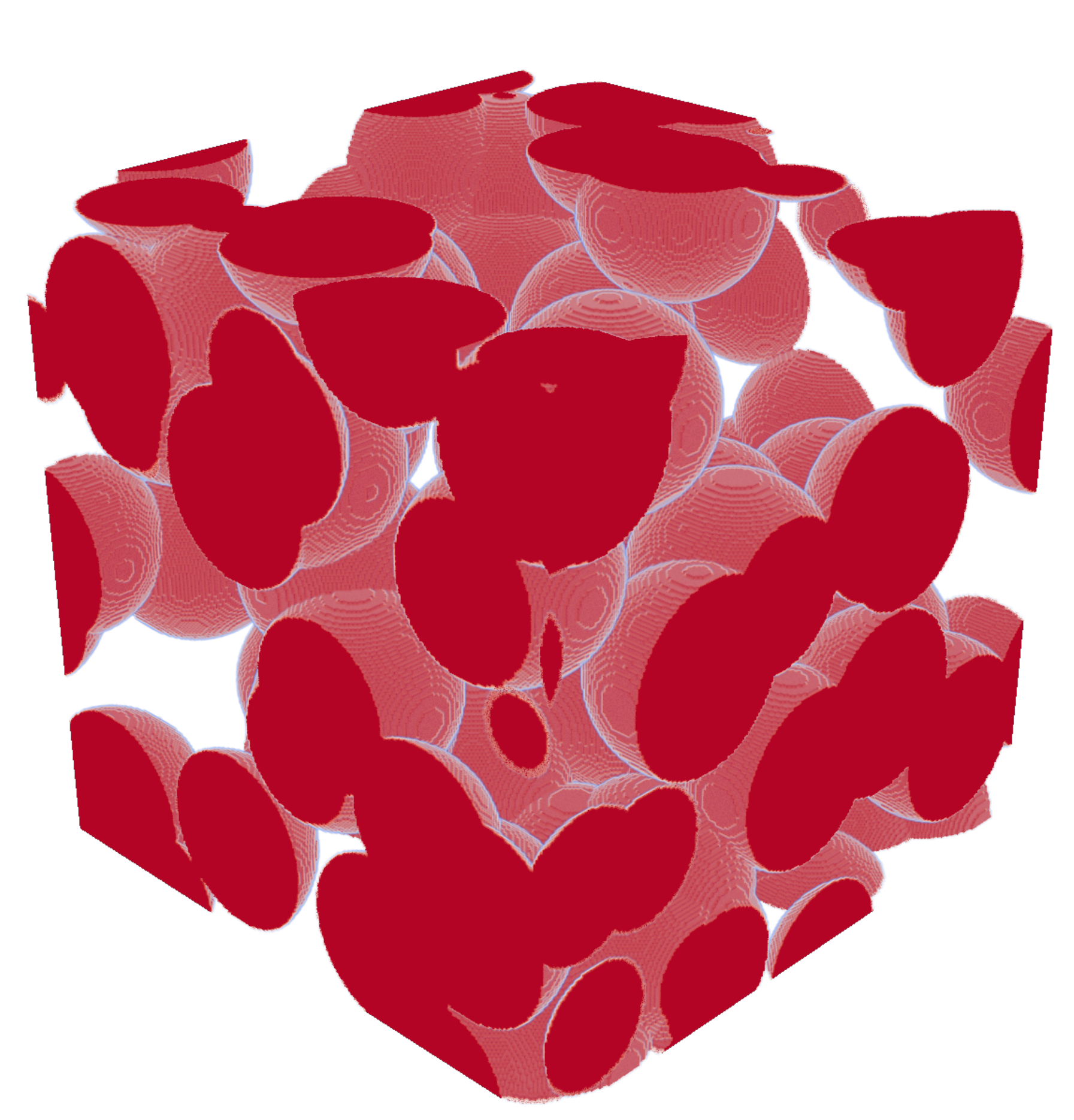}}~
    \subfloat[\label{fig:fpe}]{\includegraphics[width=0.20\textwidth]{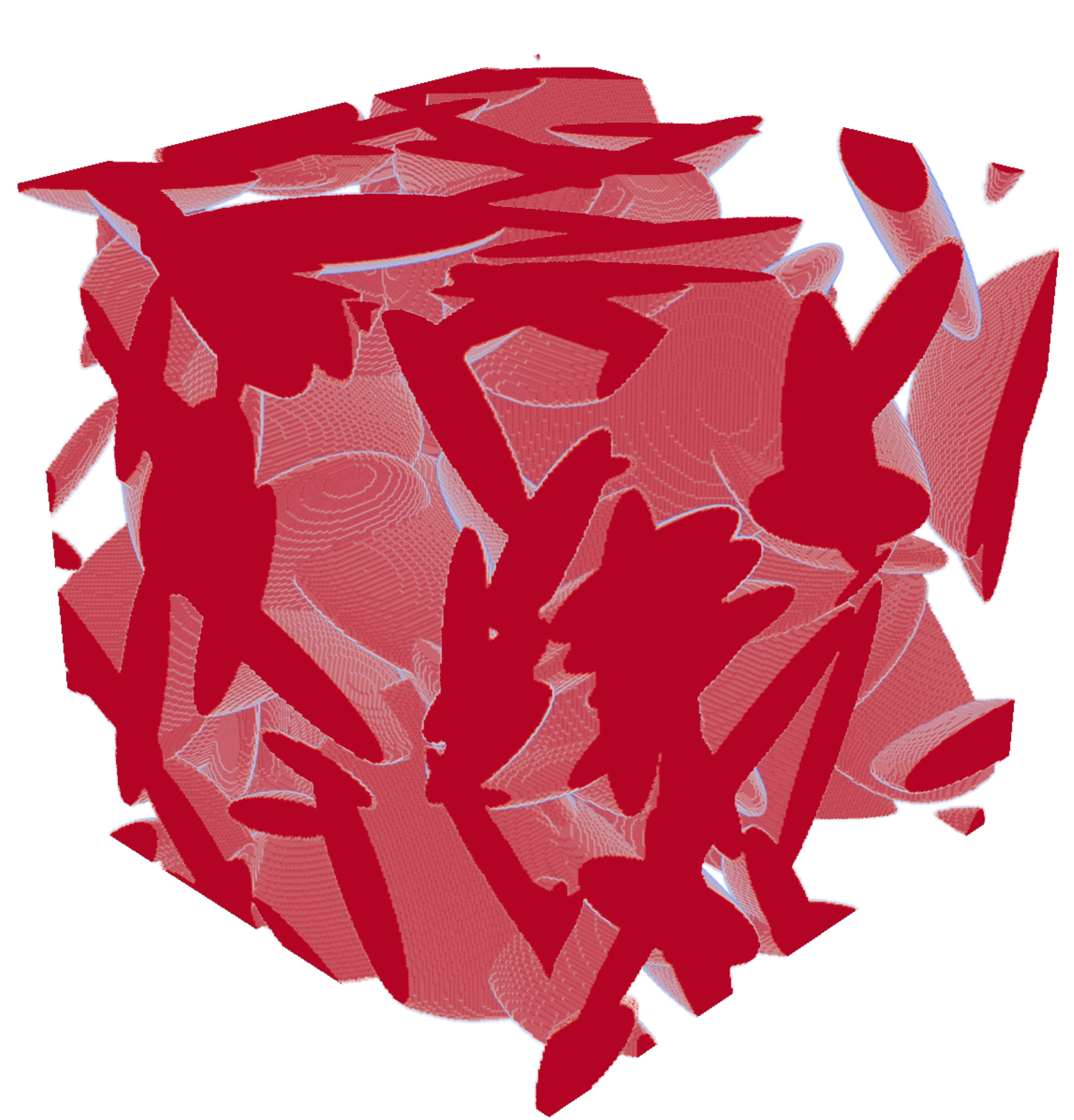}}    
    \caption{Sample images of the model highly clustered microstructures considered in this paper: (a) random checkerboard, (b) Debye random media, as well as fully penetrable (c) spheres and (d) oblate ellipsoids. The images shown in (a)-(d) correspond to $1/8$th of the total volume of the microstructures used in the calculations, and have phase 2 (red) volume fraction $\phi_2=1/2$. To facilitate visualization of the microstructure, phase 1 is transparent in these images.}
    \label{fig:models}
\end{figure}

We also consider dispersions of fully penetrable disks, spheres, as well as randomly oriented ellipses and oblate ellipsoids \cite{To02a}. Fully penetrable disk/sphere (FPD/S) systems are prepared by decorating the points of a Poisson point process with radius $a$ disks/spheres that are free to overlap with one another. The volume fraction of the \textit{matrix phase} of these systems is given by $\phi_1=\exp[-\rho v_1(a)]$, in which $\rho$ is the number density of the underlying Poisson point pattern and $v_1(a)$ is the volume of a $d$-dimensional sphere of radius $a$ \cite{To02a,To83b}. The two-point function for the matrix phase of these microstructures is given by $S_2^{(1)}=\exp[-\ln\phi_1 v_2(r;a)/v_1(a)]$, in which $v_2(r;a)$ is the union volume of two $d$-dimensional spheres of radius $a$ whose centers are a distance $r$ apart. The model fully penetrable ellipse/ellipsoid (FPE/E) systems are prepared similarly to the disk/sphere systems, except that each particle is given a random orientation in space. Their $\phi_1$ and $S_2^{(1)}$ are also mathematically analogous to those of the FPD/S systems \cite{To02a,To83b}. Note that the aforementioned systems are realistic models of consolidated media such as sandstones and sintered materials \cite{To86d}. Sample images of microstructures of these four highly clustered models in 3D with $\phi_2=1/2$ are shown in Fig. \ref{fig:models}.

By definition, a phase-inversion symmetric microstructure is one in which the $n$-point correlation functions for phases 1 and 2 obey the following relation \cite{To02a}:
\begin{equation}
    S_n^{(1)}(\mathbf{x}^n;\phi_1,\phi_2)=S_n^{(2)}(\mathbf{x}^n;\phi_2,\phi_1), \textrm{    } n\geq2\label{eqn:phaseinv}.
\end{equation}
Note that the random checkerboard and Debye microstructures have two-point correlation functions that satisfy \eqref{eqn:phaseinv} for $n=2$, and thus they possess phase-inversion symmetry at the two-point level. We show in Sec. \ref{sec:threepoint_results} that these models are also phase-inversion symmetric at the three-point level. By contrast, the two-point correlation functions for the fully penetrable particle models do not satisfy \eqref{eqn:phaseinv} and, as we will show in Sec. \ref{sec:threepoint_results}, these models are also not phase-inversion symmetric at the three-point level \footnote{We note that the self-consistent approximation for the effective conductivity \cite{To02a}, i.e., $\sigma_e = (\alpha + \sqrt{\alpha^2 + 4(d-1)\sigma_1\sigma_2})/2(d-1)$, in which $\alpha \equiv \sigma_1(d\phi_1 - 1) + \sigma_2(d\phi_2 - 1)$, is phase-inversion symmetric in the sense that it is invariant under swapping of the phase labels (i.e., $1\to2$ and $2\to1$). This property is to be distinguished from microstructure phase-inversion symmetry defined by relation \eqref{eqn:phaseinv}.}.

For subsequent calculations, we specifically consider 2D $1024\times1024$ pixel and 3D $192\times192\times192$ voxel realizations of the random checkerboard microstructures. The 2D and 3D RCB systems have $a=8$ and $a=6$ respectively, and were made using a random number generator. The Debye microstructures are the same size as the RCB systems, but have characteristic length scales $a=10$ in 2D and $a=5$ in 3D. Note that the ratio of major to minor axes $b/a=5$ for both the fully penetrable ellipse and oblate ellipsoid systems. Additional details concerning the generation of the Debye and fully penetrable particle microstructures are provided in Ref. \cite{Sk24c}. 

\subsection{Results for Three-Point Parameters}\label{sec:threepoint_results}

\begin{figure}[h]
	    \subfloat[\label{fig:pis2d_zeta}]{\includegraphics[width=0.25\textwidth]{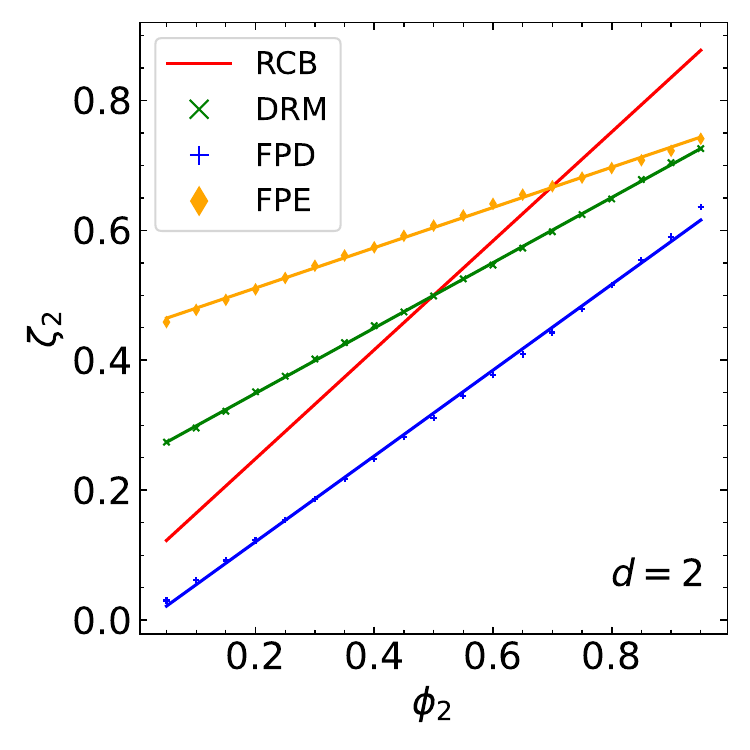}}~
	    \subfloat[\label{fig:pis3d_zeta}]{\includegraphics[width=0.25\textwidth]{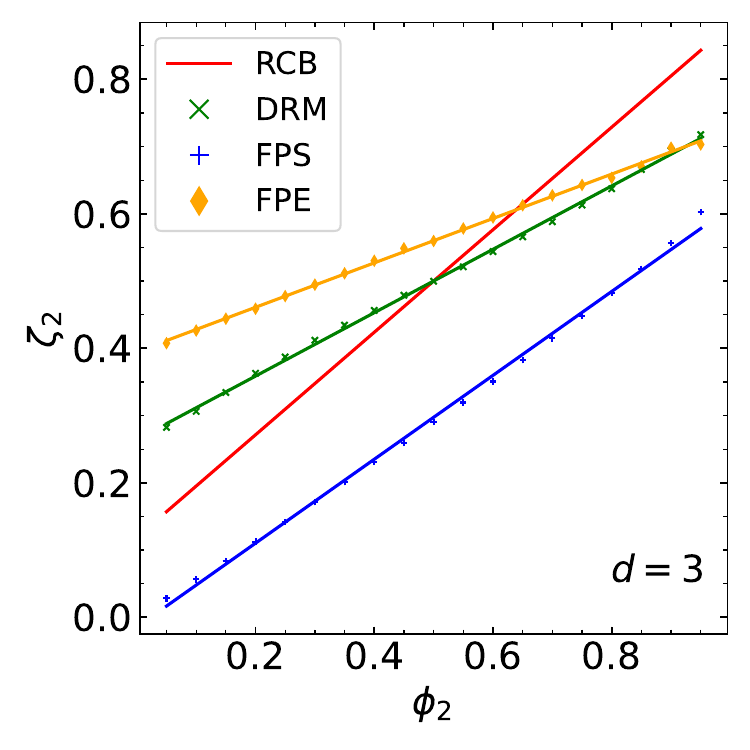}}
	\caption{Plots of the three-point parameter $\zeta_2$ as a function of phase 2 volume fraction $\phi_2$ for the model microstructures in (a) 2D and (b) 3D. The scatter plots for Debye random media as well as the fully penetrable particle models correspond to ensemble averaged values of $\zeta_2$ computed via Monte Carlo integration \cite{Sk24c}, while the corresponding solid lines are of linear regression fits (all with $R^2>0.999$) of the aforementioned numerical data. Values of $\zeta_2$ for the random checkerboards were obtained using previously calculated values of the cell-shape parameter $g$ for squares and cubes \cite{To02a}. Every value of $\zeta_2$ for a given volume fraction was averaged over 20 different configurations. Error bars for the simulated data have been omitted as they are not appreciable on the scale of this figure.}\label{fig:zeta_vs_phi}
\end{figure}

Before testing Eq. \eqref{eqn:20.84} on the aforementioned model microstructures, we first examine their values of $\zeta_2$. It is seen that the parameters $\zeta_2$ for all four models vary linearly with $\phi_2$ in the plots of $\zeta_2$ vs. $\phi_2$ in Fig. \ref{fig:zeta_vs_phi}. Note that for the penetrable particle models, we take the particles to be phase 2 and the matrix as phase 1. Additionally, we note that Debye random media have $\zeta_2=0.5\pm0.001$, implying that they must also possess phase-inversion symmetry at the three-point level. By contrast, the phase-inversion asymmetric fully penetrable particle models both have $\zeta_2\neq1/2$. 

Since the three-point parameter $\zeta_2$ for 2D and 3D Debye random media varies linearly with respect to $\phi_2$ and is equal to $1/2$ when $\phi_2=1/2$, we can fit Eq. \eqref{eqn:z2_scm} to these data in order to extrapolate ``effective" cell-shape parameters $g$ for these models. We specifically find from these fits that $g^{(2D)}_{DRM}=0.2487$ and $g^{(3D)}_{DRM}=0.2643$. While $g_{DRM}^{(2D)}$ and $g_{DRM}^{(3D)}$ are quite different from those of the random checkerboard models (i.e., $g^{(2D)}_{RCB}=0.0808$ and $g^{(3D)}_{RCB}=0.1188$ \cite{To02a}), they are respectively comparable to the $g$ values of 2D symmetric-cell materials with aspect ratio $\sim2.9$ elliptical cells, and 3D symmetric-cell materials with needle-like cells \cite{To02a}. Since Debye random media can be realized by certain space tessellations in 2D \cite{St95}, it is reasonable that this model is an ``effective" symmetric-cell microstructure. 

\section{Predictions of the self-consistent strong-contrast approximation}\label{sec:results}

In this section, we demonstrate the accurate predictive power of the self-consistent strong-contrast approximation \eqref{eqn:20.84} for the effective conductivity $\sigma_e$ across phase volume fractions and conductivity contrast ratios $\sigma_2/\sigma_1$. We also derive from \eqref{eqn:20.84} formulas for the effective conductivity in the superconducting and insulating limits, i.e., $\sigma_2/\sigma_1=\infty$ and $\sigma_2/\sigma_1=0$ respectively. Finally, we extract from these formulas accurate approximations for the percolation thresholds that incorporate three-point microstructural information via the parameter $\zeta_2$.

\subsection{Comparison to Simulations}

\begin{figure}[h]
    \subfloat[\label{fig:pis2d_se_vs_phi}]{\includegraphics[width=0.25\textwidth]{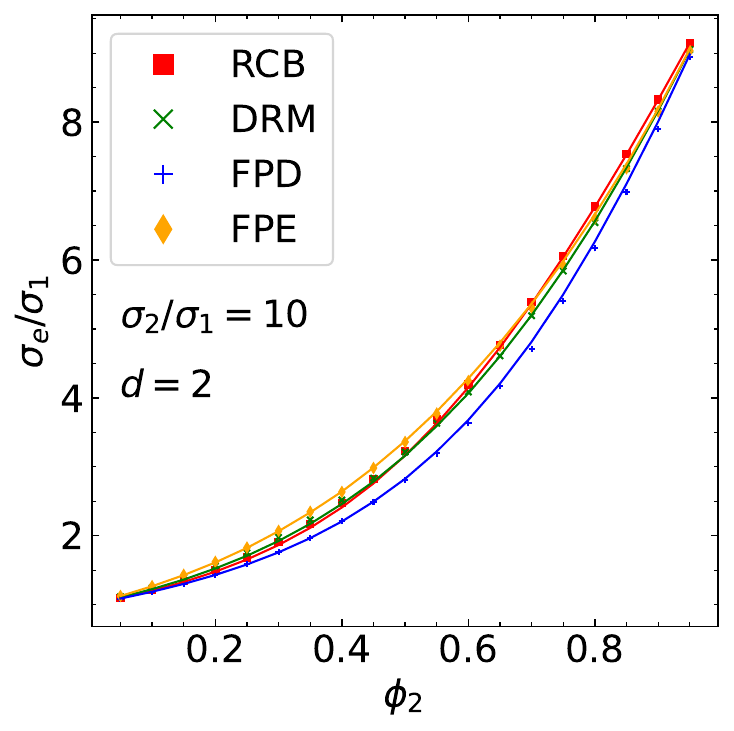}}
	\subfloat[\label{fig:pis3d_se_vs_phi}]{\includegraphics[width=0.25\textwidth]{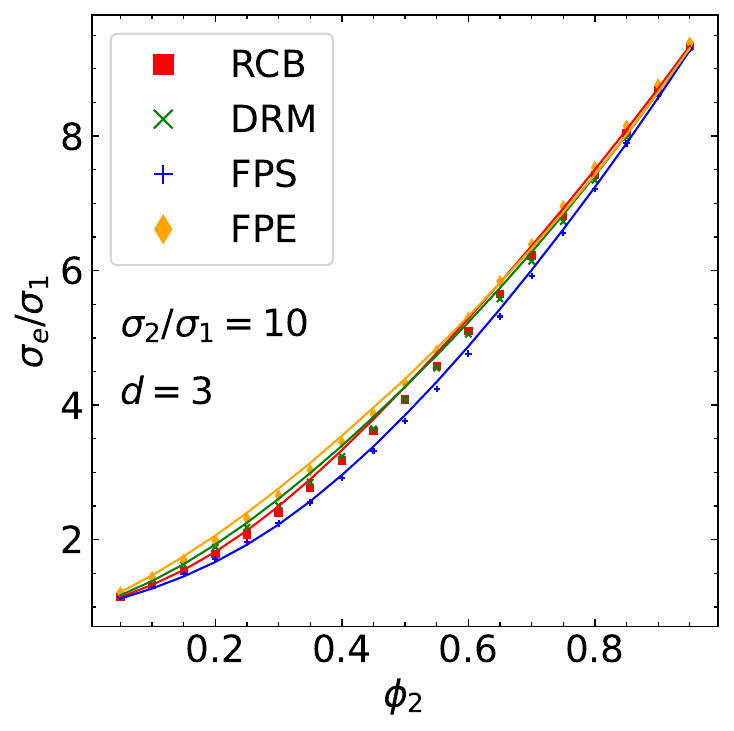}}
	\caption{Plots of the scaled effective conductivity $\sigma_e/\sigma_1$ as a function of phase 2 volume fraction $\phi_2$ for $\sigma_2/\sigma_1=10$ for the model microstructures in (a) 2D and (b) 3D. The scatter plots and solid lines correspond to simulation data and formula \eqref{eqn:20.84}, respectively. Overall, we observe excellent agreement between predictions of Eq. \eqref{eqn:20.84} and the simulated data. Each simulated value of $\sigma_e$ for a given $\phi_2$ was averaged over 20 different configurations. Error bars for the simulation data have been omitted as they are not appreciable on the scale of this figure. Simulated values of $\sigma_e$ for 2D and 3D Debye random media, random checkerboards, as well as the fully penetrable sphere and oblate ellipsoid systems were obtained via the Fast Fourier Transform (FFT)-based homogenization code described in Ref. \cite{FANS}. Those for the fully penetrable disk and ellipse models were obtained using finite element methods in COMSOL Multiphysics\textregistered.}\label{fig:se_vs_phi}
\end{figure}

For the 2D and 3D models considered here, the predictions of approximation \eqref{eqn:20.84} for the effective conductivity overall agree well with simulated values obtained across \textit{all} volume fractions for a moderate contrast ratio $\sigma_2/\sigma_1=10$, as seen in the plots of $\sigma_e$ vs. $\phi_2$ in Fig. \ref{fig:se_vs_phi}. In particular, the percent error ($100\% \times [\sigma_e^{(\textrm{theory})} - \sigma_e^{(\textrm{simulation})}]/\sigma_e^{(\textrm{simulation})}$) is between $-0.3\%$ and $3.4\%$ in 2D, and $-5.0\%$ and $1.0\%$ in 3D. The slightly larger percent error for the 3D models is attributable to certain limitations on the parameter $\zeta_2$ in approximation \eqref{eqn:20.84} for 3D systems which are detailed in Appendix \ref{app:ytransform}. In a forthcoming manuscript, we show that the self-consistent strong-contrast approximation \eqref{eqn:20.84} is accurate for Debye random media and, based on Torquato's original arguments \cite{To02a}, attribute this accuracy to the fact that this model is phase-inversion symmetric \cite{Sk24c}. We also show in that work that \eqref{eqn:20.84} is accurate for the fully penetrable particle models which we attribute to certain geometrical and topological similarities between these models and Debye random media \cite{Sk24c}. To gain a more quantitative understanding of this observed accuracy of \eqref{eqn:20.84} for these highly clustered microstructures across volume fractions, we examine in Secs. \ref{sec:other_cr} and \ref{sec:perco} its predictions for $\sigma_e$ as a function of phase conductivity contrast ratio $\sigma_2/\sigma_1$ across volume fractions.

\subsection{Theoretical Predictions Across Contrast Ratios}\label{sec:other_cr}

\begin{figure}
     \subfloat[\label{fig:pis2d_se_vs_cr}]{\includegraphics[width=0.25\textwidth]{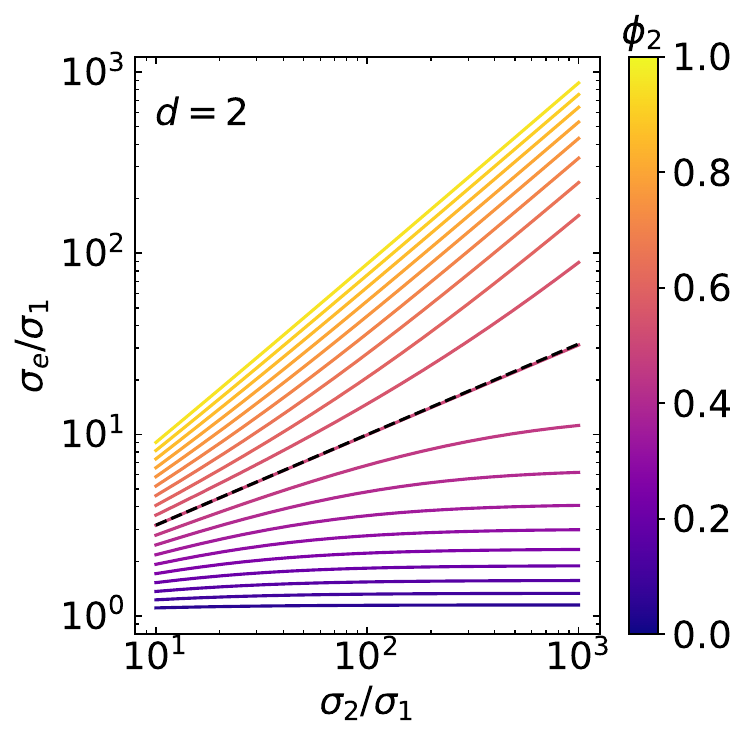}}~
	\subfloat[\label{fig:pis3d_se_vs_cr}]{\includegraphics[width=0.25\textwidth]{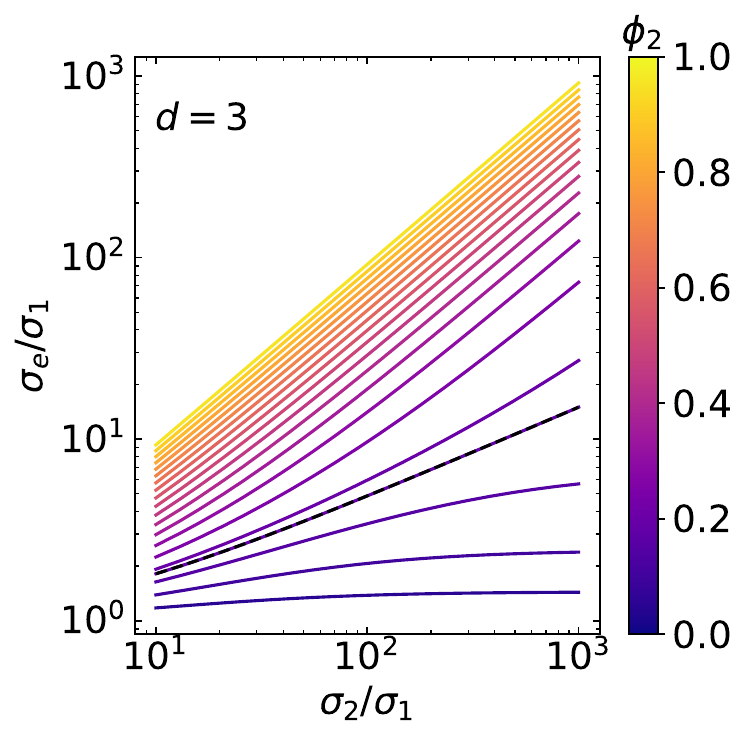}}
	\caption{Log-log plots of the scaled effective conductivity $\sigma_e/\sigma_1$ as a function of phase conductivity contrast ratio $\sigma_2/\sigma_1$ as predicted by Eq. \eqref{eqn:20.84} for Debye random media in (a) 2D and (b) 3D. Curves for different phase 2 volume fractions $\phi_2$ are color coded according to the legends on the right. In (a), the dashed black line is of Dykhne's exact formula $\sigma_e=\sqrt{\sigma_1\sigma_2}$ for 2D phase-inversion symmetric microstructures with $\phi_2=1/2$ \cite{Dy71} which we see coincides with Eq. \eqref{eqn:20.84} for $\phi_2=1/2$, $\zeta_2=1/2$, and $d=2$. In (b), the dashed black line highlights the effective conductivity of 3D Debye random media for $\phi_2=0.182$, the phase 2 percolation threshold for this model predicted by Eq. \eqref{eqn:phi2c} using the ``effective" cell-shape parameter $g_{DRM}^{(3D)}$ for this microstructure computed in Sec. \ref{sec:threepoint_results}.}\label{fig:se_vs_cr_drm}
\end{figure}

We have established the high predictive power of approximation \eqref{eqn:20.84} with numerical simulations for a fixed contrast ratio of $10$ across volume fractions for a variety of 2D and 3D models, implying its accuracy to predict values of $\sigma_e$ across all possible contrast ratios (i.e., $0\le \sigma_2/\sigma_1 \le \infty$) without relying on computationally expensive numerical simulations that become increasingly challenging when the contrast ratios become very large or very small. The accurate sensitivity of formula \eqref{eqn:20.84} to phase connectedness information of 2D and 3D microstructures across volume fractions is highlighted in the plots of $\sigma_e$ as a function of contrast ratio $\sigma_2/\sigma_1$ across volume fractions for Debye random media in Fig. \ref{fig:se_vs_cr_drm}. Specifically, it is seen that Eq. \eqref{eqn:20.84} transitions from plateauing to constant values to power-law growth in $\sigma_2/\sigma_1$ at volume fractions $\phi_2=0.5$ and $\phi_2=0.182$ in 2D and 3D, respectively. Thus, Eq. \eqref{eqn:20.84} predicts that phase 2 forms \textit{percolating clusters} at these respective volume fractions for 2D and 3D Debye random media. To investigate how Eq. \eqref{eqn:20.84} incorporates such nontrivial percolation information, we extract from it accurate analytical formulas for the thresholds $\phi_{1c}$ and $\phi_{2c}$ that depend on dimension $d$ and the parameter $\zeta_2$ in Sec. \ref{sec:perco}.

\subsection{Percolation Thresholds}\label{sec:perco}

\begin{table*}
\caption{\label{tab:thresholds}Values of the percolation thresholds $\phi_{2c}$ and $\phi_{1c}$ for the model microstructures as predicted by formulas \eqref{eqn:20.84_perco} and \eqref{eqn:20.84_perco_in}, respectively, as well as those obtained using numerical simulations. The lower limits on the thresholds for the RCB systems corresponds to site percolation when connections between nearest neighbors and next nearest neighbors are considered. By contrast, the upper limits correspond to site percolation when only nearest neighbor connections are considered. Good qualitative and/or quantitative agreement is seen between the theoretically predicted and known values of the thresholds. Note that these calculations take the matrix and particle phases of the fully penetrable particle models to be phase 1 and 2, respectively. $^{\dagger}$This value of $\phi_{1c}$ was extrapolated from values of the phase 1 percolation thresholds for fully penetrable oblate ellipsoid systems with other aspect ratios $b/a$ that were ascertained in Ref. \cite{No11}.}
\begin{ruledtabular}
\begin{tabular}{ccccc}
Model & $\phi_{2c}$, Eq. \eqref{eqn:20.84_perco} & $\phi_{2c}$, actual & $\phi_{1c}$, Eq. \eqref{eqn:20.84_perco_in} & $\phi_{1c}$, actual \\
\colrule
Random checkerboard (2D)                     & $0.500$ & $0.407 \leq \phi_{2c} \leq 0.593$ \cite{To02a} & $0.500$ & $0.407 \leq \phi_{1c} \leq 0.593$ \cite{To02a} \\
Random checkerboard (3D)                     & $0.236$ & $0.199 \leq \phi_{2c} \leq 0.312$ \cite{Lo98}  & $0.236$ & $0.199 \leq \phi_{1c} \leq 0.312$ \cite{Lo98} \\
Debye random media (2D)                      & $0.500$ & $0.500$ \cite{Ma20b,Sk21}                      & $0.500$ & $0.500$ \cite{Ma20b,Sk21} \\
Debye random media (3D)                      & $0.182$ & $0.115$ (Appendix \ref{app:drm_perco})         & $0.182$ & $0.115$ (Appendix \ref{app:drm_perco}) \\
Fully penetrable disks                       & $0.609$ & $0.676$ \cite{Qu00}                            & $0.391$ & $0.324$ \cite{Qu00} \\
Fully penetrable spheres                     & $0.337$ & $0.289$ \cite{Ri97b}                           & $0.078$ & $0.030$ \cite{Pr18} \\
Fully penetrable ellipses ($b/a=5$)          & $0.420$ & $0.460$ \cite{Xi88}                            & $0.580$ & $0.540$ \cite{Xi88} \\
Fully penetrable oblate ellipsoids ($b/a=5$) & $0.094$ & $0.176$ \cite{Ga95}                            & $0.194$ & $0.094^{\dagger}$ \cite{No11} \\
\end{tabular}
\end{ruledtabular}
\end{table*}\label{tab:phi2cvals}

By examining the behavior of \eqref{eqn:20.84} in the limit in which phase 2 becomes \textit{superconductive} relative to phase 1, we find that approximation \eqref{eqn:20.84} predicts non-trivial percolation thresholds that depend on microstructural information via the parameter $\zeta_2$. Specifically, taking the limit $\sigma_2/\sigma_1\to\infty$ in \eqref{eqn:20.84} yields
\begin{equation}
    \frac{\sigma_e}{\sigma_1} = \frac{1 + \phi_2 + (1-d + d\phi_2 - 2\phi_2)\zeta_2}{1 + \phi_2 + (1-d - 2\phi_2)\zeta_2 + d\phi_2(\zeta_2-1)}.\label{eqn:20.84_sc}
\end{equation}
Henceforth, we estimate the value of the threshold $\phi_{2c}$ from Eq. \eqref{eqn:20.84_sc} in terms of $\zeta_2$ and $d$ by setting the denominator of \eqref{eqn:20.84_sc} to zero and solving for $\phi_2$ \cite{To02a}, finding that 
\begin{equation}
    \phi_{2c} = \frac{(d-1)\zeta_2 - 1}{1 - d + (d-2)\zeta_2}\label{eqn:20.84_perco},
\end{equation}
in which $\zeta_2$ is implicitly a function of $\phi_2$. Note that there are no limitations on the value of $\zeta_2$ in Eq. \eqref{eqn:20.84_perco} in 2D, while one must have $\zeta_2\leq1/2$ in 3D. We find that Eq. \eqref{eqn:20.84_perco} predicts that \textit{all} 2D phase-inversion symmetric microstructures have threshold $\phi_{2c}=1/2$ which is consistent with observations in previous work \cite{To02a, Ma20b, Sk21}. In the case of 3D symmetric-cell materials, Eq. \eqref{eqn:20.84_perco} predicts
\begin{equation}
    \phi_{2c}(g)=\frac{-4 + 5g + \sqrt{12 - 24g + 9g^2}}{4g - 2}, \qquad g<1/2.\label{eqn:phi2c}
\end{equation}

It is also fruitful to examine the behavior of approximation \eqref{eqn:20.84} in the limit in which phase 2 becomes \textit{insulating} relative to phase 1. Taking the limit $\sigma_2/\sigma_1\to0$ in \eqref{eqn:20.84}, we find that
\begin{equation}
    \frac{\sigma_e}{\sigma_1} = \frac{1 + \phi_2 + (1-2\phi_2)\zeta_2 + d(\phi_2\zeta_2-1)}{1 + \phi_2 + (1-2\phi_1)\zeta_2 + d[\phi_2(1-\zeta_2) -1]}.\label{eqn:20.84_in}
\end{equation}
By ascertaining the value of $\phi_2$ in which approximation \eqref{eqn:20.84_in} goes to zero, we can determine when \textit{phase 1 becomes disconnected} and thus predict the phase 1 percolation threshold $\phi_{1c}$. We specifically find that
\begin{equation}
    \phi_{1c}=1-\frac{d - \zeta_2 - 1}{1 + (d-2)\zeta_2},\label{eqn:20.84_perco_in}
\end{equation}
in which $\zeta_2$ is implicitly a function of $\phi_2$. Note that there are no limits on the value of $\zeta_2$ in Eq. \eqref{eqn:20.84_perco_in} in 2D, while one must have $\zeta_2\geq1/2$ in 3D. Taken together, Eqs. \eqref{eqn:20.84_perco} and \eqref{eqn:20.84_perco_in} can be used to predict the range of volume fractions over which a microstructure is bicontinuous. For 2D systems, we find that $\phi_{2c}=1-\phi_{1c}$, implying that there is no bicontinuity of phases in this dimension. By contrast, these equations generally predict that 3D microstructures will be bicontinuous for $\phi_{2c}<\phi_2<1-\phi_{1c}$. 

\begin{figure}
     \subfloat[\label{fig:pis2d_contrast_lims}]{\includegraphics[width=0.25\textwidth]{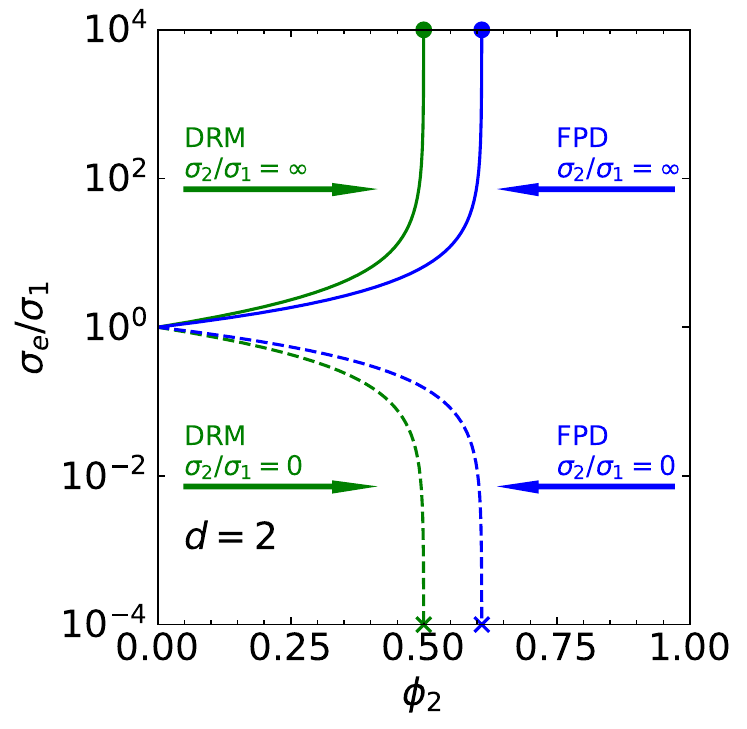}}~
	\subfloat[\label{fig:pis3d_contrast_lims}]{\includegraphics[width=0.25\textwidth]{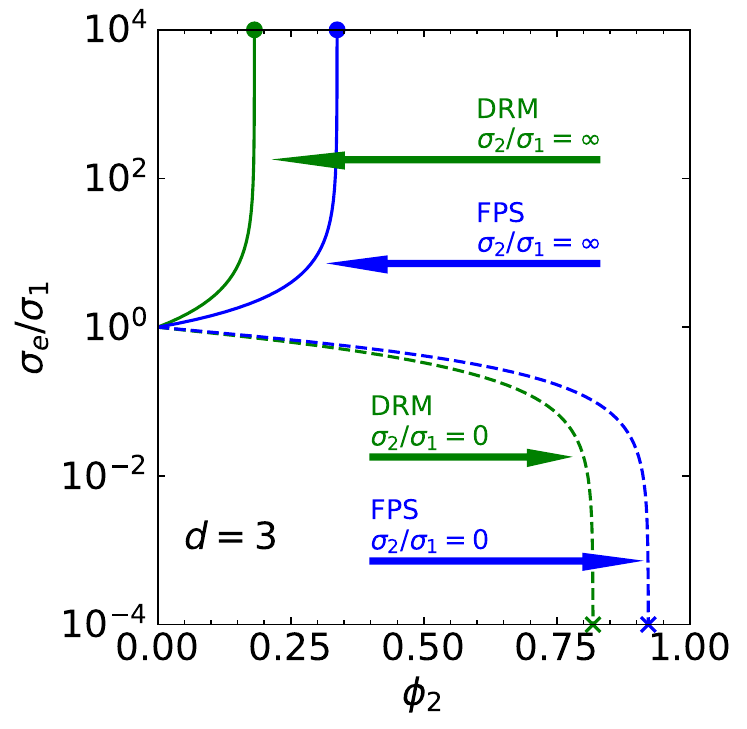}}
	\caption{Log-linear plots of the scaled effective conductivity $\sigma_e/\sigma_1$ as a function of phase 2 volume fraction $\phi_2$ as predicted by Eqs. \eqref{eqn:20.84_sc} and \eqref{eqn:20.84_in} for (a) 2D Debye random media and fully penetrable disks and (b) 3D Debye random media and fully penetrable spheres. Solid lines correspond to formula \eqref{eqn:20.84_sc} in which phase 2 is superconducting, while dashed lines correspond to formula \eqref{eqn:20.84_in} in which phase 2 is insulating. In (a) and (b), the circles and x's mark the locations of $\phi_{2c}$ predicted by Eq. \eqref{eqn:20.84_perco} and $1-\phi_{1c}$ predicted by Eq. \eqref{eqn:20.84_perco_in}, respectively, for these models.}\label{fig:se_vs_cr}
\end{figure}
\color{black}
Values of $\phi_{2c}$ and $\phi_{1c}$ for the 2D and 3D models as predicted by formulas \eqref{eqn:20.84_perco} and \eqref{eqn:20.84_perco_in}, respectively, as well as thresholds obtained via simulation methods are listed in Table \ref{tab:phi2cvals}. See Appendix \ref{app:drm_perco} for details on how we estimated the value of $\phi_{2c}$ for 3D Debye random media. We find good quantitative agreement between the predicted and numerically obtained thresholds for the phase-inversion symmetric models. Additionally, despite the fully penetrable particle models not being phase-inversion symmetric, the values of $\phi_{2c}$ and $\phi_{1c}$ predicted via formulas \eqref{eqn:20.84_perco} and \eqref{eqn:20.84_perco_in} for these systems are in good quantitative and/or qualitative agreement with the numerically obtained thresholds. Moreover, the theoretically predicted and actual intervals of volume fractions in which the 3D models are bicontinuous are in good qualitative and/or quantitative agreement with one another.

The incorporation of nontrivial phase clustering information into Eq. \eqref{eqn:20.84} across dimensions is further highlighted in the plots of this approximation in the superconducting and insulating limits (i.e., Eqs. \eqref{eqn:20.84_sc} and \eqref{eqn:20.84_in}, respectively) as functions of $\phi_2$ for 2D Debye random media and fully penetrable disks in Fig. \ref{fig:pis2d_contrast_lims} and 3D Debye random media and fully penetrable spheres in Fig. \ref{fig:pis3d_contrast_lims}. For each model microstructure, we specifically observe that Eq. \eqref{eqn:20.84_sc} diverges to infinity (i.e., the composite becomes superconductive) precisely when phase 2 forms as a sample spanning cluster at the threshold predicted by \eqref{eqn:20.84_perco}. By contrast, \eqref{eqn:20.84_in} abruptly drops to zero (i.e., the composite becomes insulating) when phase 1 becomes disconnected as predicted by Eq. \eqref{eqn:20.84_perco_in}. Finally, the nontrivial ranges of volume fractions over which these 3D microstructures are predicted to be bicontinuous are highlighted by the different volume fractions at which they are predicted to become superconducting and insulating as seen in Fig. \ref{fig:pis3d_contrast_lims}.

\section{Discussion and Conclusions}\label{sec:conclusion}

In summary, we have shown via comprehensive numerical simulations that the self-consistent strong-contrast approximation \eqref{eqn:20.84} predicts accurately the effective conductivities of four model disordered microstructures in which both phases form large, well-connected clusters in 2D and 3D across phase volume fractions. Moreover, the accuracy of Eq. \eqref{eqn:20.84} is attributable to its prediction of non-trivial percolation thresholds that incorporate $\zeta_2$ that were found to be in good quantitative and/or qualitative agreement with the known thresholds for the aforementioned models. 

One potentially fruitful area of future research is to derive an approximation analogous to Eq. \eqref{eqn:20.84} that predicts accurately the effective stiffness tensor $\mathbf{C}_e$ of phase-inversion symmetric materials. Such a derivation would likely involve the strong-contrast expansion for $\mathbf{C}_e$ derived in Ref. \cite{To97b}. It is also an outstanding problem to ascertain \textit{precise} percolation thresholds for 3D Debye random media as well as other model phase-inversion symmetric microstructures (like those examined in Ref. \cite{To22a}) in order to more comprehensively test the accuracy of $\phi_{2c}$ predicted by Eq. \eqref{eqn:phi2c}. Such calculations are challenging as they require the computationally intensive generation of many large realizations of microstructures with prescribed two-point correlation functions \cite{Ch18a,Ma20b}. 

Finally, we note that approximation \eqref{eqn:20.84} could be used to guide an inverse design problem in which microstructures with prescribed $\sigma_e$ for a given volume fraction are realized by symmetric-cell materials with targeted cell-shape parameters $g$. For simplicity, one could target tessellations of polydisperse elliptical or ellipsoidal cells whose shape parameters $g$ can be determined exactly in terms of their aspect ratios \cite{To02a}. These microstructures could be effectively realized computationally by an adaptation of Kim and Torquato's tessellation-based procedure \cite{Ki19a,Ki19b}. Finally, self-supporting microstructures, which could be manufactured by 3D printing techniques \cite{Sh15}, could be targeted in this scheme by ensuring the volume fraction of phase 2 of the designed microstructures is sufficiently above the percolation threshold as determined by Eq. \eqref{eqn:phi2c}.

\color{black}
\begin{acknowledgments}
This work was sponsored by the Army Research Office and was accomplished under Cooperative Agreement Number W911NF-22-2-0103, as well as the Air Force Office of Scientific Research Program under award No. FA9550-18-1-0514. M.S. also acknowledges the support of the Harold W. Dodds Honorific Fellowship. The authors thank the Princeton Institute for Computational Science and Engineering (PICSciE) for the computational resources.
\end{acknowledgments}

\appendix

\section{Restricting The Self-Consistent Strong-Contrast Approximation Between Three-Point Bounds}\label{app:ytransform}

We will require that the approximation \eqref{eqn:20.84} always lie between the most restrictive three-point bounds on the effective conductivity, i.e., the three-point ``Beran-type" lower and upper bounds which are given by \cite{Be65,To80,Mi81a,Si72,Sc76a,Mi82a,To02a}
\begin{equation}
    \sigma_L^{(3)} = \langle \sigma \rangle - \frac{\phi_1\phi_2(\sigma_2-\sigma_1)^2}{\langle \tilde{\sigma} \rangle + (d-1)\langle \sigma^{-1}\rangle^{-1}_{\zeta}},\label{eqn:sigma_3pt_low}
\end{equation}
and
\begin{equation}
    \sigma_U^{(3)} = \langle \sigma \rangle - \frac{\phi_1\phi_2(\sigma_2-\sigma_1)^2}{\langle \tilde{\sigma} \rangle + (d-1)\langle\sigma\rangle_{\zeta}},\label{eqn:sigma_3pt_upp}
\end{equation}
respectively. Note that $\langle \sigma \rangle=\phi_1\sigma_1 + \phi_2\sigma_2$, $\langle \tilde{\sigma}\rangle=\phi_2\sigma_1 + \phi_1\sigma_2$, and $\langle c \rangle_{\zeta}=c_1\zeta_1 + c_2\zeta_2$. By incorporating three-point microstructural information via the parameter $\zeta_2$, these bounds improve upon the well-known Hashin-Shtrikman lower and upper bounds \cite{Ha62c, Ha70}. Placing such a requirement on Eq. \eqref{eqn:20.84} will generally restrict the range of values of $\zeta_2$ beyond the normal condition that it must lie in the interval $[0,1]$ \cite{To02a}. Following Torquato \cite{To98c}, we use the so-called Y-transformation \cite{Mi92,Ch92} to determine that formula \eqref{eqn:20.84} lies between \eqref{eqn:sigma_3pt_low} and \eqref{eqn:sigma_3pt_upp} for $0\leq\zeta_2\leq1$ for 2D microstructures. Repeating the same procedure for 3D systems, we find that Eq. \eqref{eqn:20.84} lies between the three-point bounds provided that
\begin{equation}
\begin{cases}
    0 < \zeta_2 < 1/2 \textrm{ and } \phi_2 < f(\zeta_2,\gamma) \\
    1/2 < \zeta_2 < 1 \textrm{ and } \phi_2 > f(\zeta_2,\gamma),
\end{cases}\label{eqn:gen_conditions}
\end{equation}
in which $\gamma=\sigma_2/\sigma_1$ and 
\begin{multline}
    f(\zeta_2,\gamma) = \\
    \frac{\zeta_2 \left(1+2 \gamma\right)\left(2 \left(1 -  \zeta _2\right) +\left(2\zeta_2+1\right)\gamma\right)}{2 \left(\zeta_2^2-4 \zeta_2+3\right) + \left(-4\zeta _2^2+4 \zeta _2+3\right) \gamma + 2 \zeta _2 \left(\zeta _2+2\right) \gamma^2}.\label{eqn:f_function}
\end{multline}
We calculate the ranges of valid $\phi_2$ for the 3D models considered here for $\gamma=10$ via the conditions \eqref{eqn:gen_conditions} and Eq. \eqref{eqn:f_function} and list them in Table \ref{tab:ranges}. For these models, the range of $\phi_2$ values in which Eq. \eqref{eqn:20.84} violates the bounds (i.e., $1/2<\phi_2\lesssim0.93$) coincides with that in which the approximation has the largest percent error as compared to simulation, as seen in Fig. \ref{fig:pis3d_se_vs_phi} and described in Sec. \ref{sec:results}.

\begin{table}[h]
\caption{\label{tab:ranges} Ranges of $\phi_2$ values in which the approximation \eqref{eqn:20.84} for the 3D models considered here with $\gamma=10$ should lie between the three-point bounds \eqref{eqn:sigma_3pt_low} and \eqref{eqn:sigma_3pt_upp} as determined by Eq. \eqref{eqn:f_function} and constraints \eqref{eqn:gen_conditions}.}
\begin{ruledtabular}
\begin{tabular}{ccc}
3D Model&Ranges of valid $\phi_2$ values for $\gamma=10$\\
\colrule
RCB & $0<\phi_2<1/2$, $\phi_2\gtrsim0.93$ \\
DRM & $0<\phi_2<1/2$, $\phi_2\gtrsim0.85$ \\
FPS & $0<\phi_2<1/2$, $\phi_2\gtrsim0.70$ \\ 
FPE & $0<\phi_2<1/2$, $\phi_2\gtrsim0.82$ 
\end{tabular}
\end{ruledtabular}
\end{table}

\section{Estimating the Percolation Threshold of 3D Debye Random Media}\label{app:drm_perco}

\begin{figure}
     \includegraphics[width=0.45\textwidth]{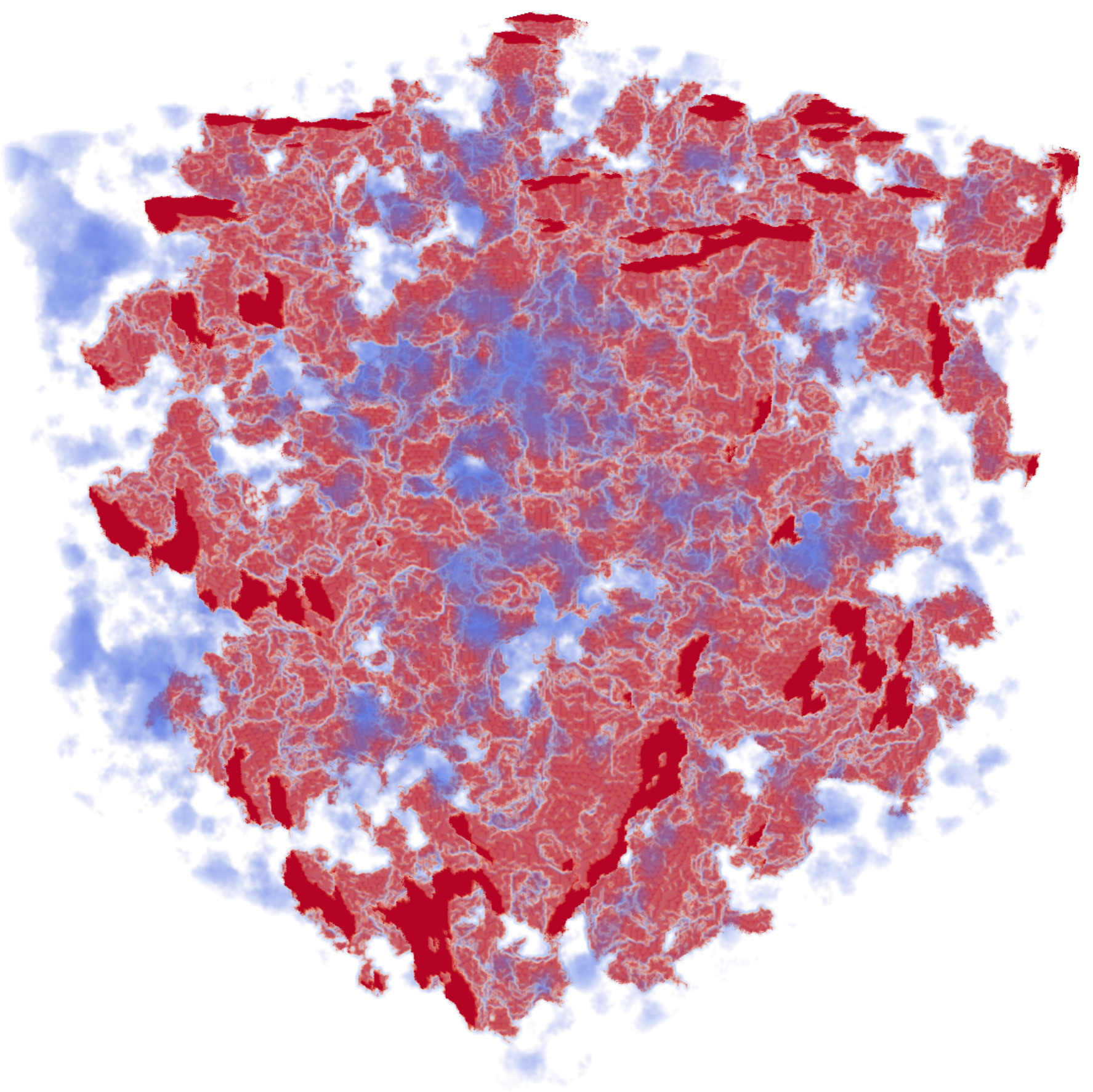}
	   \caption{A realization of 3D Debye random media at $\phi_2=0.15$ with its percolating cluster of phase 2 highlighted in red. The non-percolating domains of phase 2 are highlighted in blue, while phase 1 is transparent to facilitate visualization. The image shown here corresponds to the entire $192\times192\times192$ voxel sample.}\label{fig:drm3d_perco}
\end{figure}

In order to test the accuracy of the percolation thresholds $\phi_{2c}$ predicted via Eq. \eqref{eqn:20.84_sc} for 3D Debye random media, we estimate here the value of $\phi_{2c}$ for this model. For each volume fraction $\phi_2$, we specifically search for sample-spanning clusters in $20$ different realizations of $192\times192\times192$ voxel 3D Debye random media. To identify such clusters, we use the ``burning algorithm" \cite{St92} and consider only nearest-neighbor voxel connections for simplicity. We find that the fraction of configurations supporting percolating clusters $f_p=0$ for $\phi_2=0.08$ while $f_p=1$ for $\phi_2=0.15$, with $f_p$ increasing with $\phi_2$ for $0.08<\phi_2<0.15$. As narrowing the range of volume fractions in which $\phi_{2c}$ lies requires the computationally expensive generation of additional, larger configurations, here, we estimate $\phi_{2c}$ to be the median of this interval, i.e., $\phi_{2c}=(0.08 + 0.15)/2=0.115$. This estimated threshold is reasonably close to $\phi_{2c}=0.182$, as predicted via Eq. \eqref{eqn:20.84_sc} for this model. The percolating network of a $\phi_2=0.15$ realization of 3D Debye random media is highlighted in Fig. \ref{fig:drm3d_perco}.

%

 \newcommand{\noop}[1]{}

\end{document}